\documentclass[%
 reprint,
superscriptaddress,
 amsmath,
 amssymb,
 aps,
prc,
]{revtex4-2}

\usepackage{hyperref}
\usepackage{amsmath}
\usepackage{amssymb}
\usepackage{graphicx}
\usepackage{mathrsfs}
\usepackage{color}
\usepackage{bm}
\usepackage{times}
\usepackage{ulem}

\DeclareRobustCommand{\erase}{\bgroup\markoverwith{\textcolor{red}{\rule[.5ex]{2pt}{1.pt}}}\ULon}

\newcommand{\blue}[1]{\textcolor{blue}{{#1}}}

\newcommand{\magenta}[1]{\textcolor{magenta}{{#1}}}

\begin{document}

\title{Application of the optimized-basis generator coordinate method \\ to low-lying excited states of sd-shell nuclei
}

\author{Moemi Matsumoto}
\altaffiliation {Current affiliation: Hitachi, Ltd., 1-280, Higashi-Koigakubo, Kokubunji-shi, Tokyo 185-8601, Japan}
\affiliation{Department of Physics, Tohoku University, Sendai, 980-8578, Japan}
\author{Yusuke Tanimura}
\affiliation{Department of Physics and Origin of Matter and Evolution of Galaxies (OMEG) Institute, Soongsil University, Seoul 06978, Korea}

\author{Kouichi Hagino}
\affiliation{Department of Physics, Kyoto University, Kyoto, 606-8502, Japan}
\affiliation{ 
RIKEN Nishina Center for Accelerator-based Science, RIKEN, Wako 351-0198, Japan
}
\date{\today}

\begin{abstract}
We apply the optimized-basis generator coordinate method (OptGCM) to sd-shell nuclei, 
$^{20}$Ne, $^{24}$Mg, and $^{28}$Si. 
This method 
variationally optimizes 
both the basis Slater determinants in the generator coordinate method (GCM) 
and the corresponding weight coefficients. 
To analyze the low-lying excited states of 
those nuclei, we implement the angular momentum 
projection. 
With the Skyrme interaction, we show that 
the simultaneous optimzation of the basis functions and the weight factors lowers the energy of the excited states 
and at the same time leads to an appreciable effect on transition 
probabilities. 
These results highlight the effectiveness of the OptGCM method.

\end{abstract}


\maketitle

\section{Introduction}
Collective motion is a characteristic feature of quantum many-body systems. 
In atomic nuclei, it has been known that various types of collective motions appear, such as the rotational motion of deformed nuclei and surface vibrations of spherical nuclei. 
The amplitude of those motions are often large,  beyond the quadratic approximation around the ground state. Such large-amplitude 
collective motions 
play a crucial role in several phenomena in atomic nuclei, such as nuclear fission and shape coexistence.

To develop microscopic theories for collective motions is one of the central challenges in nuclear physics.
Among such theories, the generator coordinate method (GCM) has been widely employed as a beyond-mean-field approach to large-amplitude collective motions 
\cite{RingSchuck,Bender03,Niksic11,YoGoBe19}.
In this method, a many-body state is constructed as a linear combination of configurations generated along specified collective coordinates, that characterize the essential degrees of freedom of the dynamics.
The weight coefficients of the linear superposition are determined  variationally.
This method has been 
extensively used in nuclear structure theory, after implementing restoration of broken symmetries with the parity and 
the angular momentum projections\cite{Yao2013,Yao2013-2,Yao2015,Yao2015-2,Bender2003,Bender2003-2,Bally2014,Duguet2003,Rodriguez2008,Rodriguez2002,Rodriguez2004,Rodriguez2007,Robledo2019,Rodriguez2012}.

A potential drawback of this method 
is that the selection of collective coordinates often relies on empirical knowledge.
For instance, multipole moments of atomic nuclei have 
often been employed for such collective coordinates, 
but these may not be sufficient to accurately describe a given system. 
This problem represents a longstanding and significant challenge inherent to the method \cite{RingSchuck}.
As a matter of fact, 
recent studies have demonstrated that the 
conventional empirical approaches to select collective coordinates may be insufficient to efficiently 
capture collective dynamics \cite{Hizawa21,Hizawa22,Kumar23}. 
It is thus desirable to develop frameworks that do not rely on pre-fixed collective coordinates, 
in order not to miss essential degrees of freedom.
Towards this direction, 
a stochastic selection of the basis has been considered in 
Refs. \cite{Itagaki03,Descouvemont18,Descouvemont20,Shinohara06,Fukuoka13}.  Methods based on configuration interaction 
have also been proposed in Refs. \cite{Faessler69,Satpathy70,Satpathy70-erratum,Pillet17,Robin16,Robin17}.

Recently, 
we have developed a variational extension of the GCM (the optimized-basis GCM: OptGCM) that allows simultaneous optimization of both the single-particle states of the basis Slater determinants (SDs) and their corresponding weight coefficients \cite{Matsumoto23}. 
See also Ref. \cite{Shimizu12} and Refs. \cite{Myo23,Hasegawa20} for similar 
attempts in the frameworks of the Monte-Carlo shell model (MCSM) \cite{Shimizu12} and antisymmetrized molecular dynamics (AMD) \cite{Myo23,Hasegawa20}, respectively. 
In our previous work \cite{Matsumoto23, Matsumoto24}, we applied the OptGCM to the intrinsic ground states of $^{16}$O and $^{28}$Si, and demonstrated that 
the basis optimization leads to a better description 
of the ground state of the systems.

In this paper, 
we extend our previous study and 
apply the OptGCM to study low-lying excited states of atomic nuclei \cite{Matsumoto25}. 
To this end, we implement the angular momentum 
projection and compute spectra and transition probabilities 
of {\it sd}-shell nuclei. 

The paper is organized as follows. 
In Sec. II, we detail the formalism of the OptGCM with 
angular momentum projection. 
In Sec. III, we apply the OptGCM to 
$^{20}$Ne, $^{24}$Mg, and $^{28}$Si 
and discuss the role of basis optimization. 
We then summarize the paper in Sec. IV 
with future perspectives.

\section{Method}
\subsection{Optimized-basis GCM (OptGCM)}\label{ssec:OptGCM}

In the GCM method, the trial wave function is expressed as 
\begin{equation}
|\Psi\rangle = \sum_{a=1}^M f_a|\Phi_a\rangle,
\label{eq:trial_gcm}
\end{equation}
where $|\Phi_a\rangle$ are SDs constructed as antisymmetrized products of $A$ orthonormal single-particle orbitals $\varphi^{(a)}_i$ ($i=1,\ldots,A$), 
$A$ being the mass number of a nucleus.
The total energy is given by
\begin{equation}
E = \frac{\langle\Psi|H|\Psi\rangle}{\langle\Psi|\Psi\rangle}
= \frac{\sum_{ab} f_a^* f_b H_{ab}}{\sum_{ab} f_a^* f_b N_{ab}},
\label{eq:Eintr}
\end{equation}
where $N_{ab} = \langle\Phi_a|\Phi_b\rangle$ and 
$H_{ab} = \langle\Phi_a|H|\Phi_b\rangle$ denote the norm and Hamiltonian kernels, respectively.

In the usual GCM, the energy is minimized with respect 
only to the weight coefficients $f_a$ for a pre-fixed 
Slater determinants, $|\Phi_a\rangle$. 
In contrast, In the 
OptGCM, the total energy 
is minimized with respect both to the weight coefficients $f_a$ and the single-particle orbitals $\varphi^{(a)}_i$ \cite{Matsumoto23}. That is, we impose the conditions of
\begin{equation}
    \cfrac{\partial E}{\partial f_a^*}=0,
    \label{eq-f}
\end{equation}
and
\begin{equation}
\cfrac{\delta E}{\delta\varphi_i^{(a)*}}=0. 
 \label{eq-swf}
\end{equation}
Note that 
Eq. (\ref{eq-f}) yields the Hill-Wheeler equation, 
which is coupled to the equation for the single-particle 
wave functions, Eq. (\ref{eq-swf}).
We carry out the optimization with the conjugate gradient method \cite{NumericalRecipes}, with 
initial SDs constructed with Woods-Saxon potentials with various quadrupole deformations. For the initial weight coefficients, 
we take $f_a = 1$ for all $a$.

In this work, we adopt the SIII parameter set of the Skyrme energy density functional \cite{Beiner75}, and expand the single-particle wave functions on an axial harmonic oscillator basis \cite{Vautherin73} with 14 major shells. 
We impose axial and reflection symmetries, 
while we neglect the time-odd terms in the functional. 
The pairing correlations are also omitted for simplicity.

\subsection{Angular momentum projection}

To directly compare theoretical results 
with experimental data, it is essential to restore rotational symmetry through the angular momentum projection. 
Ideally, the basis SDs in the OptGCM should be variationally optimized for each angular momentum $I$. 
Since this procedure is computationally demanding,  
we, as a first step, adopt a simplified approach, in which the basis SDs are optimized without the angular momentum projection.
That is, we take the variation before the projection (VBP) for the angular momentum projection, rather than 
the variation after the projection (VAP). 
To this end, 
we first optimize the basis SDs $|\Phi_a\rangle$ and the weight coefficients $f_a$ for 
the intrinsic ground states, as described in the previous 
subsection. 
After the optimum set of the basis SDs are so obtained, 
we perform the angular momentum projection to restore the rotational symmetry and calculate low-lying excited states. 
That is, we apply the projection operator 
$\hat P_{MK}^{I}$ onto Eq. (\ref{eq:trial_gcm}) and obtain 
\begin{align}
|\Psi_{IM}\rangle
&=
\sum_{a,K}f_{aK}\hat P_{MK}^I|\Phi_a\rangle.
\end{align}
Here, 
$I$ is the total angular momentum and $M$ and $K$ are its 
projection onto the $z$-axis in the 
laboratory and the body-fixed frames, respectively. 
The weights are then re-determined by solving the Hill-Wheeler equation with the projected SDs for each 
angular momentum, $I$. 

With the total wave function so obtained, 
we calculate the reduced electric transition probabilities,  $B(E2)$. 
The $B(E\lambda)$ for a transition between a state with spin $I_i$ and that with $I_f$ is given by \cite{RingSchuck}
\begin{equation}
    B(E\lambda, I_i \rightarrow I_f) = \frac{1}{2I_i + 1} \left|\langle I_f \| \hat{Q}^{(e)}_\lambda \| I_i \rangle\right|^2,
    \label{eq:BE}
\end{equation}
where $\langle I_f \| \hat{Q}^{(e)}_\lambda \| I_i \rangle$ is the reduced matrix element, and $\hat{Q}_{\lambda\mu}^{(e)}$ 
is  
the electric multipole operator given by
\begin{equation}
    \hat{Q}_{\lambda\mu}^{(e)} = e \int d{\bm r}\, \hat{\rho}_p({\bm r}) r^\lambda Y_{\lambda\mu}(\hat{\bm r}). 
\end{equation}
Here, $\hat{\rho}_p$ is the proton density operator and $Y_{\lambda\mu}$ is the spherical harmonic.

For comparison, we also perform conventional GCM calculations with the quadrupole moment 
for the collective coordinate.  
In this approach, the basis SDs \( |\Phi(Q_2^{(a)})\rangle \) are local ground states obtained by the constrained Hartree-Fock (CHF) method with the constraint
\begin{eqnarray}
   \langle\Phi_a(Q_2^{(a)})|\hat Q_{2} | \Phi_a(Q_2^{(a)})\rangle &=& Q_2^{(a)}.
   \end{eqnarray}
Here, $\hat{Q}_2$ is the quadrupole operator, 
   \begin{eqnarray}
 \hat{Q}_2&=&\int d{\bm r} \  \hat \rho(\bm{r}) r^2 Y_{20}(\hat{\bm{r}}),
\end{eqnarray}
where $\hat\rho$ and $Y_{20}$ 
are the density operator and the spherical harmonic, respectively.
As in the OptGCM, the weights are determined 
after the angular momentum projection is performed.

\section{Results and discussions}

\subsection{Intrinsic ground states}

Let us now apply the OptGCM to the $^{20}$Ne, $^{24}$Mg, and $^{28}$Si nuclei. 
For $^{20}$Ne and $^{28}$Si, we superpose 10 SDs, 
while we superpose 8 SDs for 
$^{24}$Mg in both OptGCM and GCM calculations.
We have confirmed that 
the results are not significantly changed even if we increase the number of SDs. 

We first discuss the intrinsic ground state. 
Table \ref{tab:intrinsic} summarizes the energies of
the intrinsic ground state of
$^{20}$Ne, $^{24}$Mg, and $^{28}$Si obtained with the Hartree-Fock (HF), the GCM, and the OptGCM calculations.
In the case of $^{20}$Ne, the energy decreases by 0.56~MeV from HF to GCM, and 
it further decreases by 0.87~MeV from GCM to OptGCM. 
Similarly, for $^{24}$Mg ($^{28}$Si), the energy decreases by 
0.74 (1.19)~MeV from HF to GCM, and by 0.91 (0.78)~MeV from GCM to OptGCM. 
These are consistent with the results of our previous study presented in Ref. \cite{Matsumoto23}.

\begin{table}[]
  \centering
  \caption{The energies (in MeV) of the intrinsic ground states of
  $^{20}$Ne, $^{24}$Mg, and $^{28}$Si obtained with the HF, GCM, and OptGCM calculations.}
\begin{ruledtabular}
\begin{tabular}{lccc}
& HF & GCM & OptGCM \\ \hline
$^{20}$Ne& $-153.28$ & $-153.84$ & $-154.71$ \\ 
$^{24}$Mg& $-191.21$ & $-191.95$ & $-192.86$ \\ 
$^{28}$Si& $-228.74$ & $-229.93$ & $-230.71$ \\ 
\end{tabular}
\end{ruledtabular}
\label{tab:intrinsic}
\end{table}

\begin{figure*}
\includegraphics[width=16cm]{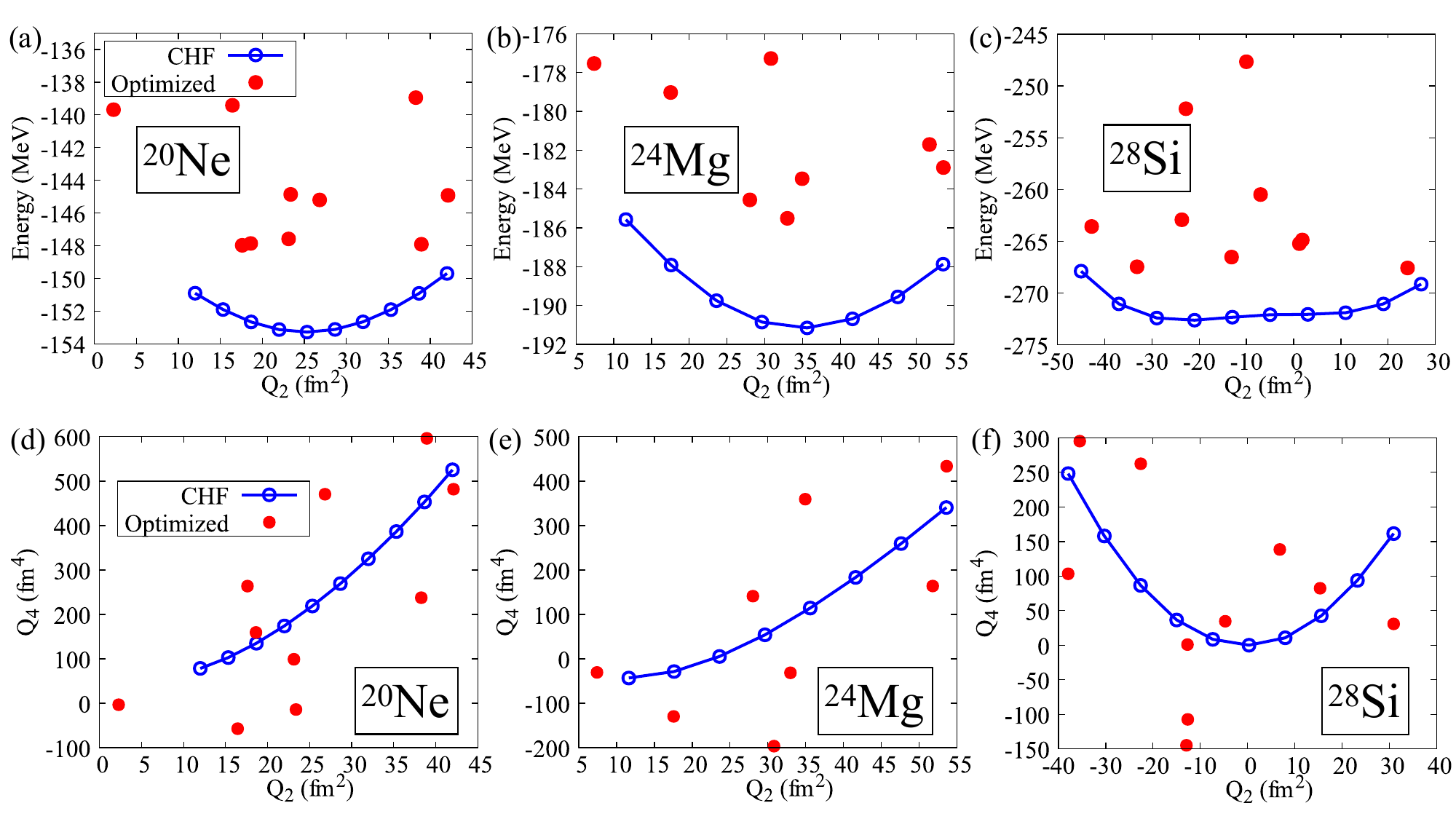}
\caption{ 
The upper row: 
the energy expectation values of the basis SDs obtained with the CHF (the blue circles) and the OptGCM (the red dots) 
for (a) $^{20}$Ne, (b) $^{24}$Mg, and (c) $^{28}$Si nuclei. 
These are plotted as a function of the expectation values of the quadrupole moments $Q_2$.  
The lower row: 
the quadrupole and the hexadecapole moments, $Q_2$ and $Q_4$, of the basis SDs 
obtained with the CHF (the blue circles) and the OptGCM (the red dots)
for (d) $^{20}$Ne, (e) $^{24}$Mg, and (f) $^{28}$Si nuclei. 
}
\label{fig:basis}
\end{figure*}

To understand the energy gains due to the basis optimization in OptGCM, Figure \ref{fig:basis} shows several quantities obtained with the optimized basis states 
for the intrinsic ground state. 
Figures \ref{fig:basis} (a), (b), and (c) present the expectation value of the energy for 
each basis state as a function of \( Q_2 \) for the \(^{20}\)Ne, \(^{24}\)Mg, and \(^{28}\)Si nuclei, respectively. 
The blue lines represent the potential energy curves (PECs) obtained using CHF calculations. 
Since we do not take into account the pairing correlation in the present calculations, 
the CHF calculations for 
\(^{20}\)Ne and \(^{24}\)Mg
do not converge around \( Q_2 = 0 \) due to level crossings. 
We exclude those points in the PEC. 

In all the three nuclei studied in this paper, 
the optimized basis states lie upper than each of the PECs, thus correspond
to excited states for a given value of $Q_2$. 
This is similar to our previous result for \(^{16}\)O \cite{Matsumoto23}. 
Despite the large diagonal elements of the Hamiltonian in the optimized basis, 
the ground-state energies obtained with OptGCM are still lower than those with the conventional GCM. 
This energy gain arises from the enhanced off-diagonal matrix elements in the optimized basis \cite{Matsumoto24}. 
These results indicate that taking a superposition of solely the local ground states 
does not necessarily lead to a good ground state.  
The OptGCM incorporates many-body correlations in a more complex way than 
the conventional GCM.

Figures \ref{fig:basis} (d), (e), and (f) present scatter plots of the quadrupole moment \( Q_2 \) and hexadecapole moment \( Q_4 \), calculated for each basis state in \(^{20}\)Ne, \(^{24}\)Mg, and \(^{28}\)Si, respectively. 
The red dots represent the results with the OptGCM basis states, while the blue circles 
correspond to those obtained from the CHF calculations with constraint on \( Q_2 \).  
While the local ground states obtained with CHF trace the valley of the potential energy surface, 
the OptGCM results exhibit considerable fluctuations around the valley. 
This behavior indicates that collective dynamics of $Q_4$ as well as $Q_2$ are efficiently  
taken into account in the OptGCM states. 
These results suggest that, at least for the nuclei considered in this study, 
it is necessary to include higher-order multipole moments, at least up to \( Q_4 \), in the collective coordinates in order to optimize the intrinsic ground states. 
Furthermore, we point out that the OptGCM is also advantageous from a computational 
perspective, as it reduces the technical efforts required to perform multidimensional GCM calculations, 
which would require many mesh points in a multi-dimensional potential energy surface. 

\begin{figure*}
\includegraphics[width=16cm]{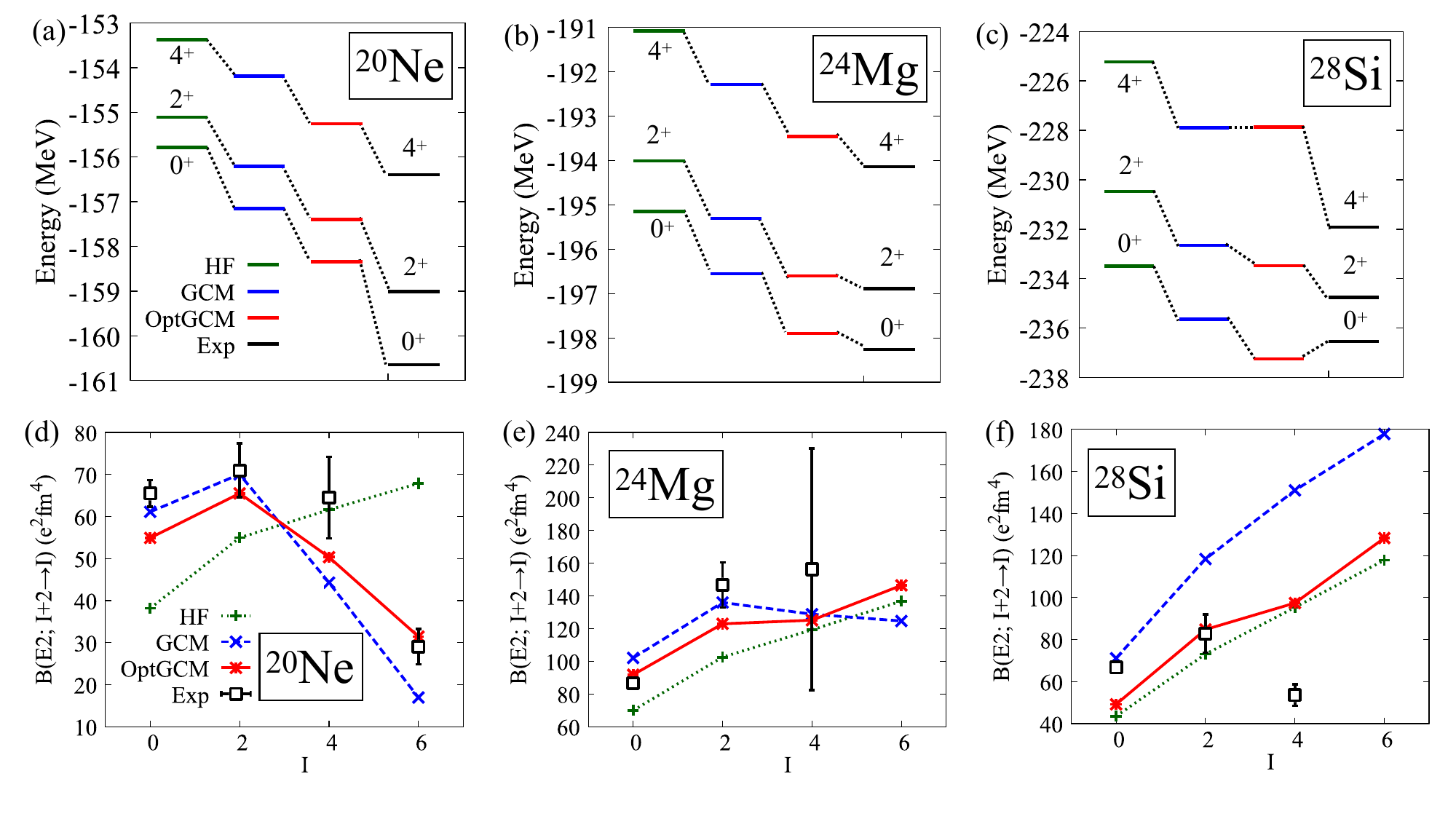}
\caption{ 
The energy spectra 
for the (a) $^{20}$Ne, (b) $^{24}$Mg, and (c) $^{28}$Si nuclei 
obtained with the HF, the GCM, and the OptGCM with the angular momentum projection, 
in comparison with the experimental data.  
 Comparisons of the calaulated $B(E2)$ values with the experimental data 
are also shown in the lower panels. 
The experimental data are taken from Ref. \cite{Tilley98}. 
}
\label{fig:energy_be2}
\end{figure*}

\subsection{Low-lying states with angular momentum projection}

We next discuss the results for excited states. 
Figures \ref{fig:energy_be2} (a), (b), and (c) show the calculated and experimental ground-state rotational spectra of the \(^{20}\)Ne, \(^{24}\)Mg, and \(^{28}\)Si nuclei, respectively, obtained with the HF, the GCM, and the OptGCM calculations. 
The experimental data are taken from Ref. \cite{Tilley98}.
One can see that the energies of the \(0^+\), \(2^+\), and \(4^+\) states decrease successively from the HF to the GCM 
and further to the OptGCM, similar to the intrinsic states shown in Table \ref{tab:intrinsic}.  
Therefore, for these low-lying states, the OptGCM results represent an improvement over 
both the HF and the standard GCM.
Notice that the decrease of the ground-state energies after the 
angular momentum projection is comparable to that for the intrinsic states.

As the angular momentum increases, the differences in energy between the OptGCM and the GCM become smaller.
This is the case particularly for the 4$^+$ state in $^{28}$Si. 
A part of the reason for this is 
that the basis states of the OptGCM in these calculations are optimized with respect to the intrinsic ground states and it is
unreasonable to assume that such basis states can efficiently describe excited states with large excitation energies. 
This limitation would be removed by optimizing basis states for each angular momentum individually.
We also note that the OptGCM yields the ground state energy of $^{28}$Si that is lower than the experimental value. 
This, however, would not be a serious problem given that the SIII parameter set of the Skyrme interaction 
is designed for HF calculations, rather than beyond-mean-field calculations such as those with the GCM and the OptGCM. 
Even with such interaction, 
one would still be able to discuss differences of the results among the models.  

Figures \ref{fig:energy_be2} (d), (e), and (f) present the \( B(E2) \) transition strengths for 
the $^{20}$Ne, $^{24}$Mg, and $^{28}$Si nuclei. 
For all the three nuclei, both the GCM and the OptGCM lead to a 
reasonably good agreement with the experimental data, except for 
$B(E2:6^+\to4^+)$ in $^{28}$Si. 
In particular, in the case of $^{20}$Ne, 
the angular momentum dependence of \( B(E2) \) 
is well described both with the GCM and the OptGCM, whereas the HF leads to an inconsistent angular momentum dependence. 
The reduction in \( B(E2) \) at $I=2$ is known as the effect of band termination, signaling the end of the 
ground-state rotational band \cite{Afanasjev99}.
Table \ref{tab:ne20_be2} compares the \( B(E2) \) values of $^{20}$Ne obtained in our calculations (the OptGCM, the GCM, and the HF) 
with the results of other theoretical approaches, such as the 
relativistic Hartree-Bogoliubov (RHB) \cite{Marevic18}, the relativistic Hartree plus BCS (RH+BCS) \cite{Zhou16}, and 
the antisymmetrized molecular dynamics (AMD) \cite{Kimura04}, as well as with the experimental data \cite{Tilley98}. 
The results of the OptGCM and the GCM are consistent with those of the other beyond-mean-field methods 
and show good agreement with the experimental data.
We mention that 
those beyond-mean-field methods do not impose reflection symmetry, in contrast to our calculations. However,   
the potential energy surfaces of $^{20}$Ne reported in Refs. \cite{Zhou16,Marevic18} suggest that the impact of 
reflection symmetry breaking is negligible.

\begin{table*}[]
  \centering
  \caption{The observed (Exp.) and the calculated (OptGCM, GCM, and HF) intraband $E2$ transition probabilities $B(E2;I_i^\pi\rightarrow I_f^\pi)$, in units of e$^2$fm$^4$, within the $K^\pi=0_1^+$ band for $^{20}$Ne. 
  These are compared with the results of 
  the relativistic Hartree-Bogoliubov (RHB) with DD-PC1 interaction \cite{Marevic18}, 
  the relativistic Hartree+BCS (RH+BCS) with PC-PK1 interaction \cite{Zhou16},
  the deformed-basis AMD with Gogny D1S interaction \cite{Kimura04}, 
  and the $(sd)^4$ shell model (SM) \cite{Tomoda78}.}
\begin{ruledtabular}
\begin{tabular}{lcccccccc}
$I_i^\pi\rightarrow I_f^\pi$ & Exp.       & OptGCM  & GCM  & HF   &RHB & RH+BCS & AMD  & SM   \\ \hline
$2^+\rightarrow 0^+$         & 65$\pm$3  & 54.9 & 61.1 & 38.1 &54  & 74     & 70.3 & 57   \\
$4^+\rightarrow 2^+$         & 71$\pm$6  & 65.5 & 70   & 55   &89  & 71     & 83.7 & 69.9 \\
$6^+\rightarrow 4^+$         & 64$\pm$10 & 50.3 & 44.3 & 61.6 &85  & 68     & 52.7 & 57.9 \\
$8^+\rightarrow 6^+$         & 29$\pm$4  & 31.4 & 16.9 & 67.9 &     &        & 21   & 35.5 \\ 
\end{tabular}
\end{ruledtabular}
\label{tab:ne20_be2}
\end{table*}


\section{Summary and future perspectives}

We have investigated the low-lying states of the \(^{20}\)Ne, \(^{24}\)Mg, and \(^{28}\)Si nuclei 
using the OptGCM developed in our previous work \cite{Matsumoto23}. 
In contrast to the previous work, we have implemented the angular momentum 
projection in the OptGCM to compute a spectrum and transition probabilities. 
To reduce a computational cost, we have first optimized the basis states for the intrinsic ground state and then performed 
the angular momentum projection. 
We have compared the results with those obtained by the HF, the conventional GCM, and the available experimental data. 
Comparisons were also made with a few previous theoretical studies.

Although some variations specific to each nucleus were observed, we have observed several common features. 
First, the OptGCM consistently yields lower energies for the low-lying states compared to the HF and the GCM. 
According to the variational principle, a lower energy implies a closer approximation to the exact solution for a given 
many-body Hamiltonian. 
This indicates that the OptGCM leads to an improvement over the HF and the GCM, at least for the nuclei studied 
in this paper. 
Second, an analysis of the energies and multipole moments of the basis SDs 
suggests that nontrivial many-body correlations are incorporated into the OptGCM wave function 
as a result of the basis optimization.

The present method can be further extended in several ways. 
This includes 
optimizing the basis states for each angular momentum, as has been done in MCSM \cite{Shimizu12}.  
This will improve the description of excited states. 
An extension of the framework to include the pairing correlation 
by replacing the SDs by Hartree-Fock-Bogoliubov (HFB) states is also an important future work, 
particularly for describing open-shell nuclei as well as nuclear deformation. 
Furthermore, removing the constraints of axial and reflection symmetry 
will allow a description of more general nuclear shapes. 
Systematic applications to a variety of nuclei, along with comparisons with other theoretical approaches such as the conventional GCM, the self-consistent collective coordinate (SCC) method \cite{Marumori80,Matsuo86,Matsuo00,Hinohara08}, and the dynamical GCM with collective momenta \cite{Goeke80,Hizawa21,Hizawa22}, will also help to identify the relevant collective degrees of freedom.

Another interesting future direction is to extend the current method to time-dependent frameworks 
in order to describe nuclear dynamics, such as fusion and fission processes. 
Preliminary calculations using the AMD basis have suggested that a description of many-body quantum tunneling 
\cite{Hasegawa20} will be possible with such an extension. 
In this context, it is worth noting that several research groups have recently developed time-dependent extensions of the GCM framework \cite{Li23,Li24,Li2025,Marevic24}.

\section*{Acknowledgments}
This work was supported by JST SPRING grant No. JPMJSP2114, and the JSPS KAKENHI grants No. 19K03861, 23K03414, and 24KJ0352. 
M. M. acknowledges the support from the Graduate Program on Physics for the Universe (GP-PU) of Tohoku University. 
Y. T. acknowledges support from the Basic Science Research Program of the National Research Foundation of Korea (NRF) under grants No. RS-2024-00361003, RS-2024-00460031, and RS-
2021-NR060129.

\bibliography{optgcm_amproj}

\end{document}